\documentclass[iop]{emulateapj}
\usepackage{lineno}

\shorttitle{An r-Process Enhanced Star in Tucana~III}
\shortauthors{Hansen et al.}

\begin{document}

\title{An R-process enhanced star in the dwarf galaxy Tucana III\altaffilmark{*}}

\altaffiltext{*}{This paper includes data gathered with the 6.5 meter
  Magellan Telescopes located at Las Campanas Observatory, Chile.}

\author{
T.~T.~Hansen\altaffilmark{1},
J.~D.~Simon\altaffilmark{1},
J.~L.~Marshall\altaffilmark{2},
T.~S.~Li\altaffilmark{3,2},
D.~Carollo\altaffilmark{4,5},
D.~L.~DePoy\altaffilmark{2},
D.~Q. Nagasawa\altaffilmark{2},
R.~A.~Bernstein\altaffilmark{1},
A.~Drlica-Wagner\altaffilmark{3},
F.~B.~Abdalla\altaffilmark{6,7},
S.~Allam\altaffilmark{3},
J.~Annis\altaffilmark{3},
K.~Bechtol\altaffilmark{8},
A.~Benoit-L{\'e}vy\altaffilmark{9,6,10},
D.~Brooks\altaffilmark{6},
E.~Buckley-Geer\altaffilmark{3},
A. Carnero Rosell\altaffilmark{11,12},
M.~Carrasco~Kind\altaffilmark{13,14},
J.~Carretero\altaffilmark{15,16},
C.~E.~Cunha\altaffilmark{17},
L.~N.~da Costa\altaffilmark{11,12},
S.~Desai\altaffilmark{18},
T.~F.~Eifler\altaffilmark{19},
A.~Fausti Neto\altaffilmark{11},
B.~Flaugher\altaffilmark{3},
J.~Frieman\altaffilmark{3,20},
J.~Garc\'ia-Bellido\altaffilmark{21},
E.~Gaztanaga\altaffilmark{15},
D.~W.~Gerdes\altaffilmark{22},
D.~Gruen\altaffilmark{17,23},
R.~A.~Gruendl\altaffilmark{13,14},
J.~Gschwend\altaffilmark{11,12},
G.~Gutierrez\altaffilmark{3},
D.~J.~James\altaffilmark{24,25},
E.~Krause\altaffilmark{17},
K.~Kuehn\altaffilmark{26},
N.~Kuropatkin\altaffilmark{3},
O.~Lahav\altaffilmark{6},
R.~Miquel\altaffilmark{27,16},
A.~A.~Plazas\altaffilmark{19},
A.~K.~Romer\altaffilmark{28},
E.~Sanchez\altaffilmark{29},
B.~Santiago\altaffilmark{30,11},
V.~Scarpine\altaffilmark{3},
R.~C.~Smith\altaffilmark{25},
M.~Soares-Santos\altaffilmark{3},
F.~Sobreira\altaffilmark{11,31},
E.~Suchyta\altaffilmark{32},
M.~E.~C.~Swanson\altaffilmark{14},
G.~Tarle\altaffilmark{22},
A.~R.~Walker\altaffilmark{25}
\\ \vspace{0.2cm} (DES Collaboration) \\
}

\altaffiltext{1}{Observatories of the Carnegie Institution of Washington, 813 Santa Barbara St., Pasadena, CA 91101, USA}
\altaffiltext{2}{George P. and Cynthia Woods Mitchell Institute for Fundamental Physics and Astronomy, and Department of Physics and Astronomy, Texas A\&M University, College Station, TX 77843,  USA}
\altaffiltext{3}{Fermi National Accelerator Laboratory, P. O. Box 500, Batavia, IL 60510, USA}
\altaffiltext{4}{INAF- Osservatorio Astronomico di Torino - Strada Osservatorio 20, Pino Torinese, 10020, Italy}
\altaffiltext{5}{ARC Centre of Excellence for All-Sky Astrophysics (CAASTRO)}
\altaffiltext{6}{Department of Physics \& Astronomy, University College London, Gower Street, London, WC1E 6BT, UK}
\altaffiltext{7}{Department of Physics and Electronics, Rhodes University, PO Box 94, Grahamstown, 6140, South Africa}
\altaffiltext{8}{LSST, 933 North Cherry Avenue, Tucson, AZ 85721, USA}
\altaffiltext{9}{CNRS, UMR 7095, Institut d'Astrophysique de Paris, F-75014, Paris, France}
\altaffiltext{10}{Sorbonne Universit\'es, UPMC Univ Paris 06, UMR 7095, Institut d'Astrophysique de Paris, F-75014, Paris, France}
\altaffiltext{11}{Laborat\'orio Interinstitucional de e-Astronomia - LIneA, Rua Gal. Jos\'e Cristino 77, Rio de Janeiro, RJ - 20921-400, Brazil}
\altaffiltext{12}{Observat\'orio Nacional, Rua Gal. Jos\'e Cristino 77, Rio de Janeiro, RJ - 20921-400, Brazil}
\altaffiltext{13}{Department of Astronomy, University of Illinois, 1002 W. Green Street, Urbana, IL 61801, USA}
\altaffiltext{14}{National Center for Supercomputing Applications, 1205 West Clark St., Urbana, IL 61801, USA}
\altaffiltext{15}{Institut de Ci\`encies de l'Espai, IEEC-CSIC, Campus UAB, Carrer de Can Magrans, s/n,  08193 Bellaterra, Barcelona, Spain}
\altaffiltext{16}{Institut de F\'{\i}sica d'Altes Energies (IFAE), The Barcelona Institute of Science and Technology, Campus UAB, 08193 Bellaterra (Barcelona) Spain}
\altaffiltext{17}{Kavli Institute for Particle Astrophysics \& Cosmology, P. O. Box 2450, Stanford University, Stanford, CA 94305, USA}
\altaffiltext{18}{Department of Physics, IIT Hyderabad, Kandi, Telangana 502285, India}
\altaffiltext{19}{Jet Propulsion Laboratory, California Institute of Technology, 4800 Oak Grove Dr., Pasadena, CA 91109, USA}
\altaffiltext{20}{Kavli Institute for Cosmological Physics, University of Chicago, Chicago, IL 60637, USA}
\altaffiltext{21}{Instituto de Fisica Teorica UAM/CSIC, Universidad Autonoma de Madrid, 28049 Madrid, Spain}
\altaffiltext{22}{Department of Physics, University of Michigan, Ann Arbor, MI 48109, USA}
\altaffiltext{23}{SLAC National Accelerator Laboratory, Menlo Park, CA 94025, USA}
\altaffiltext{24}{Astronomy Department, University of Washington, Box 351580, Seattle, WA 98195, USA}
\altaffiltext{25}{Cerro Tololo Inter-American Observatory, National Optical Astronomy Observatory, Casilla 603, La Serena, Chile}
\altaffiltext{26}{Australian Astronomical Observatory, North Ryde, NSW 2113, Australia}
\altaffiltext{27}{Instituci\'o Catalana de Recerca i Estudis Avan\c{c}ats, E-08010 Barcelona, Spain}
\altaffiltext{28}{Department of Physics and Astronomy, Pevensey Building, University of Sussex, Brighton, BN1 9QH, UK}
\altaffiltext{29}{Centro de Investigaciones Energ\'eticas, Medioambientales y Tecnol\'ogicas (CIEMAT), Madrid, Spain}
\altaffiltext{30}{Instituto de F\'\i sica, UFRGS, Caixa Postal 15051, Porto Alegre, RS - 91501-970, Brazil}
\altaffiltext{31}{Universidade Federal do ABC, Centro de Ci\^encias Naturais e Humanas, Av. dos Estados, 5001, Santo Andr\'e, SP, Brazil, 09210-580}
\altaffiltext{32}{Computer Science and Mathematics Division, Oak Ridge National Laboratory, Oak Ridge, TN 37831}

\begin{abstract}
Chemically peculiar stars in dwarf galaxies provide a window for exploring the birth environment of stars with varying chemical enrichment. We present a chemical abundance analysis of the brightest star in the newly discovered ultra-faint dwarf galaxy candidate Tucana III. Because it is particularly bright for a star in an ultra-faint Milky Way satellite, we are able to measure the abundance of 28 elements, including 13 neutron-capture species. This star, DES~J235532.66$-$593114.9 (DES~J235532), shows a mild enhancement in neutron-capture elements associated with the $r$-process and can be classified as an $r$-I star. DES~J235532 is the first $r$-I star to be discovered in an ultra-faint satellite, and Tuc~III is the second extremely low-luminosity system found to contain $r$-process enriched material, after Reticulum~II. Comparison of the abundance pattern of DES~J235532 with $r$-I and $r$-II stars found in other dwarf galaxies and in the Milky Way halo suggests a common astrophysical origin for the neutron-capture elements seen in all $r$-process enhanced stars. We explore both internal and external scenarios for the $r$-process enrichment of Tuc~III and show that with abundance patterns for additional stars it should be possible to distinguish between them. 
\end{abstract}

\keywords{}

\section{Introduction} \label{sec:intro}

A small fraction ($\sim$5\%) of metal-poor stars in the Milky Way (MW) halo show overabundances of elements produced by the rapid neutron-capture process \citep{barklem2005}. These so-called $r$-process enhanced stars can be divided into two groups depending on their level of $r$-process enhancement: the $r$-I stars with $0.3 \lesssim \mathrm{[Eu/Fe]} \lesssim 1.0$ and the $r$-II stars with $\mathrm{[Eu/Fe]} > 1.0$ \citep{beers2005}. It has long been recognized that the abundance pattern for the heavy neutron-capture elements in $r$-II stars follows the solar system $r$-process pattern \citep[e.g.,][]{mcwilliam1995,hill2002,sneden2003}. The same was recently found to be true for the moderately enhanced and more numerous $r$-I stars \citep{siqueira2014}. 

It has been suggested that $r$-process enhanced stars are the result of mass transfer from a binary companion \citep{QW01}. However, long term radial velocity monitoring performed by \citet{hansen2011,hansen2015} of 17 $r$-process enhanced stars over a period of eight years showed that these stars have a binary frequency of $\sim$18\%, consistent with the binary frequency for ordinary metal-poor field giants \citep{carney2003}. Since many stars with $r$-process enhancements are therefore not in binaries, the unusual abundance patterns of the $r$-I and $r$-II stars must be the result of nucleosynthesis events in the early Universe. The suggested astrophysical sites for production of $r$-process elements include neutron star mergers (NSMs) \citep[e.g.,][]{lattimer74,meyer89,freiburghaus99,metzger10,goriely11,korobkin12,bauswein13,rosswog14} and various types of supernovae (SNe), including magneto-rotationally driven SNe (also known as jet SNe) \citep{cameron2003,fujimoto08,winteler2012} and neutrino-driven winds in core-collapse SNe \citep[ and references therein]{arcones2013}. The NSMs and jet SNe are referred to as main $r$-process sites and can produce the heavy $r$-process elements. The yields of the ejected material from these sites are predicted to match the Solar System's heavy $r$-process-element pattern \citep{fujimoto08,bauswein13}. Nucleosynthesis in neutrino driven winds in core-collapse SNe is referred to as the weak $r$-process and is thought to produce the lighter $r$-process elements up to $A \sim 100$ \citep{arcones07}.

In addition to their presence in the halo, a number of $r$-process enhanced stars have also been found in dwarf galaxies that are satellites of the Milky Way, highlighted by the discovery that seven of the nine brightest stars in the ultra-faint dwarf Reticulum~II (Ret~II) are $r$-II stars \citep{ji2016,roederer16}. Several long-studied dwarfs such as Draco, Ursa Minor (UMi), Sculptor, Fornax, and Carina also contain $r$-process enhanced stars \citep[e.g.,][]{cohen2010,tsujimoto2015,lemasle2014,shetrone2003}, although the implications of these stars have received relatively little attention in the literature. These dwarf galaxies, including both the ultra-faint dwarfs and the classical dwarf spheroidals, provide a unique environment for exploring nucleosynthesis events in the early universe. Unlike stars from the Milky Way halo that may have been accreted by the Galaxy from a variety of smaller systems, the stars in dwarf galaxies are more likely to have remained in their birthplaces \citep{griffen16}.  Hence, not only can the chemical abundances of the stars trace just a few nucleosynthesis events, but the galaxy as a whole also preserves a record of the environment in which the events occurred.

A large population of new Milky Way satellites, many of them quite nearby, has been identified over the past two years \citep[e.g.,][]{bechtol2015,koposov2015,drlica-wagner2015}, led by discoveries from the Dark Energy Survey \citep[DES;][]{des2016,flaugher2015}.  One of the closest of these objects is Tucana~III, located just 25~kpc away \citep{drlica-wagner2015}.  \citet{simon2016} presented spectroscopic confirmation of Tuc~III and identified several bright member stars,
including one star at $V=15.8$ that is the brightest known star in an ultra-faint dwarf galaxy. This star, DES~J235532.66$-$593114.9 (hereafter shortened to DES~J235532), provides a unique opportunity to measure the abundance of heavy elements that ordinarily cannot be detected in the spectra of individual stars beyond the Milky Way.

In this paper, we present high-resolution Magellan/MIKE spectroscopy of DES~J235532 and analyze its chemical abundance pattern.  In \S\ref{observations} we briefly discuss our observations and data reduction.  We describe our abundance analysis in \S\ref{analysis}.  In \S\ref{results} we compile a comparison sample of $r$-process enhanced stars in dwarf galaxies and in the Milky Way and examine the similarities between them. We consider the implications of these results for the chemical enrichment of Tuc~III and $r$-process nucleosynthesis in dwarf galaxies in \S\ref{discussion}, and we present our conclusions in \S\ref{conclusion}.

\section{Observations}
\label{observations}

\begin{figure}
\begin{center}
\includegraphics[scale=0.7]{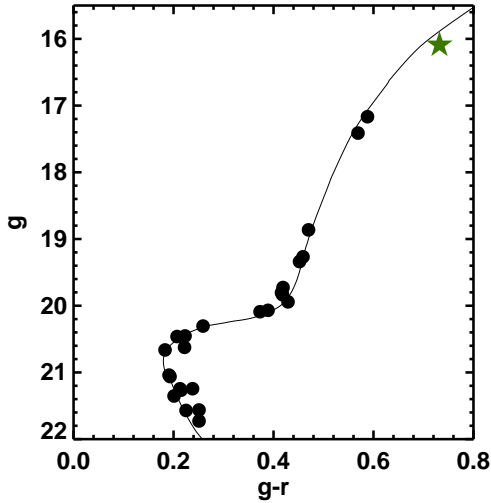}
\end{center}
\caption{DES color-magnitude diagram for members of Tucana~III from \citet{simon2016}. A Dartmouth isochrone \citep{dotter2008} for an age of 13~Gyr, $\mathrm{[Fe/H]} = -2.5$  and $\mathrm{[\alpha/Fe]} = 0.2$ is overplotted (gray curve). DES~J235532 is marked with a green star. \label{figcm}} 
\end{figure}

\citet{simon2016} used medium-resolution spectroscopy to identify a sample of 26 member stars in Tuc~III, including ten stars on the red giant branch that are brighter than $g\sim20$. The color-magnitude diagram of Tuc~III is displayed in Fig~\ref{figcm}, and DES~J235532 (shown as the green star) was selected as the brightest member of the system. The $g$ and $r$ magnitudes currently available from DES are calibrated at the ~2\% level and have statistical uncertainties ranging from 0.001 to 0.02~mag \citep{drlica-wagner2015,simon2016}. 

We observed DES~J235532 with the MIKE spectrograph \citep{bernstein2003} on the Magellan/Clay telescope on the night of 2015 August 14. We obtained a total of 3.5~hr of integration time in generally poor seeing ($>1\arcsec$), with some light cirrus present. The observations were made with a $1\arcsec \times 5\arcsec$ slit, providing a spectral resolution of $R=28000$ on the blue channel and $R=22000$ on the red channel. We reduced the data with the latest version of the Carnegie MIKE pipeline originally described by
\citet{kelson2003}.\footnote{http://code.obs.carnegiescience.edu/mike}
Because DES~J235532 is so bright, the combined spectrum is of higher quality than most high-resolution spectra of stars in dwarf galaxies, reaching a S/N of $\sim$20 per pixel at 4100~\AA, and $\sim$50 per pixel at 5500~\AA.

\section{Stellar Parameter Measurements and Abundance Analysis}
\label{analysis}

Stellar parameters for DES~J235532 have been determined from equivalent width measurements of 120 Fe~I and 25 Fe~II lines listed in Table~\ref{tab1}. The effective temperature ($T_{\rm eff}$) was derived from excitation equilibrium and then corrected for the offset between spectroscopic and photometric temperature scales using the method of \citet{frebel2013}. The surface gravity was determined by ensuring abundance agreement between Fe~I and Fe~II lines, and the microturbulence was calculated by eliminating any abundance trend with reduced equivalent width for both Fe~I and Fe~II lines. The final parameters are listed in Table \ref{tab2}. By varying the temperature and evaluating the resultant trend of the abundances with excitation potential we estimate the uncertainty of $T_{\rm eff}$ to be 100~K. The standard deviation of the Fe~I abundances is 0.18~dex, calculated from the 120 Fe~I lines used for parameter determination. Based on this value and the uncertainty in $T_{\rm eff}$ we estimate uncertainties of 0.3~dex for log$g$ and 0.3 km~s$^{-1}$ for the microturbulence. We have used a spectrum of Arcturus to check our line list and find good agreement with the parameters derived by \cite{ramirez2013}. The velocity of DES~J235532 was measured by cross-correlating each order of the spectrum with a high S/N MIKE spectrum of the bright, metal-poor giant HD~122563, assuming a velocity of $v_{{\rm hel}} = -26.51$~km~s$^{-1}$ for HD~122563 \citep{chubak12}. The MIKE velocity of DES~J235532 is in good agreement with the average velocity for the star of $v_{hel} = -103.2 \pm 1.2$~km~s$^{-1}$ measured by \citet{simon2016} from two medium-resolution spectra, as well as with the mean radial velocity of $v_{hel} = -102.4$~km~s$^{-1}$ found by \citeauthor{simon2016} for 26 member stars of Tuc~III.

\begin{deluxetable}{lrrrrr}
\tablecaption{Fe~I and Fe~II lines used for parameter determination.\label{tab1}}
\tablewidth{0pt}
\tabletypesize{\scriptsize}
\tablehead{\colhead{Species} & \colhead{$\lambda$} & \colhead{$\chi$} & \colhead{$\log$ $gf$} & \colhead{EW} & \colhead{$\log \epsilon$}\\
\colhead{} &\colhead{(\AA)} & \colhead{(eV)} & \colhead{(dex)} & \colhead{(m\AA)} & \colhead{(dex)}}
\startdata
Fe~I & 3743.36  &    0.99  &  -0.785  &   215.5 &  5.29\\ 
Fe~I & 3787.88  &    1.01  &  -0.859  &   177.6 &  5.03\\ 
Fe~I & 3849.97  &    1.01  &  -0.871  &   202.6 &  5.27\\ 
Fe~I & 3865.52  &    1.01  &  -0.982  &   163.1 &  4.93\\ 
Fe~I & 3917.18  &    0.99  &  -2.155  &   121.6 &  5.24\\ 
Fe~I & 3949.95  &    2.18  &  -1.251  &    98.0 &  5.21\\ 
Fe~I & 4076.63  &    3.21  &  -0.529  &    64.3 &  4.99\\ 
Fe~I & 4114.44  &    2.83  &  -1.303  &    52.2 &  5.11\\ 
Fe~I & 4132.06  &    1.61  &  -0.675  &   170.0 &  5.32\\ 
Fe~I & 4132.90  &    2.85  &  -1.006  &    57.8 &  4.92\\ 
.    &  .   	&    .	   &  	.     &	   .	&  .    \\
.    & 	.	&    .	   &  	.     &	   .	&  .   \\
.    & 	.	&    .	   &  	.     &	   .	&  .  \\
Fe~II & 4416.82 &    2.78  &  -2.410  &    61.2 & 4.93 \\ 
Fe~II & 4489.18 &    2.83  &  -2.971  &    34.6 & 5.09 \\
Fe~II & 4508.28 &    2.86  &  -2.250  &    71.9 & 5.04 \\ 
Fe~II & 4515.33 &    2.84  &  -2.450  &    63.2 & 5.07 \\ 
\enddata
\tablecomments{
The complete version of Table~\ref{tab1} is available online only.
A short version is shown here to
illustrate its form and content.
}

\end{deluxetable}

\begin{figure*}
\begin{center}
\includegraphics[scale=0.28]{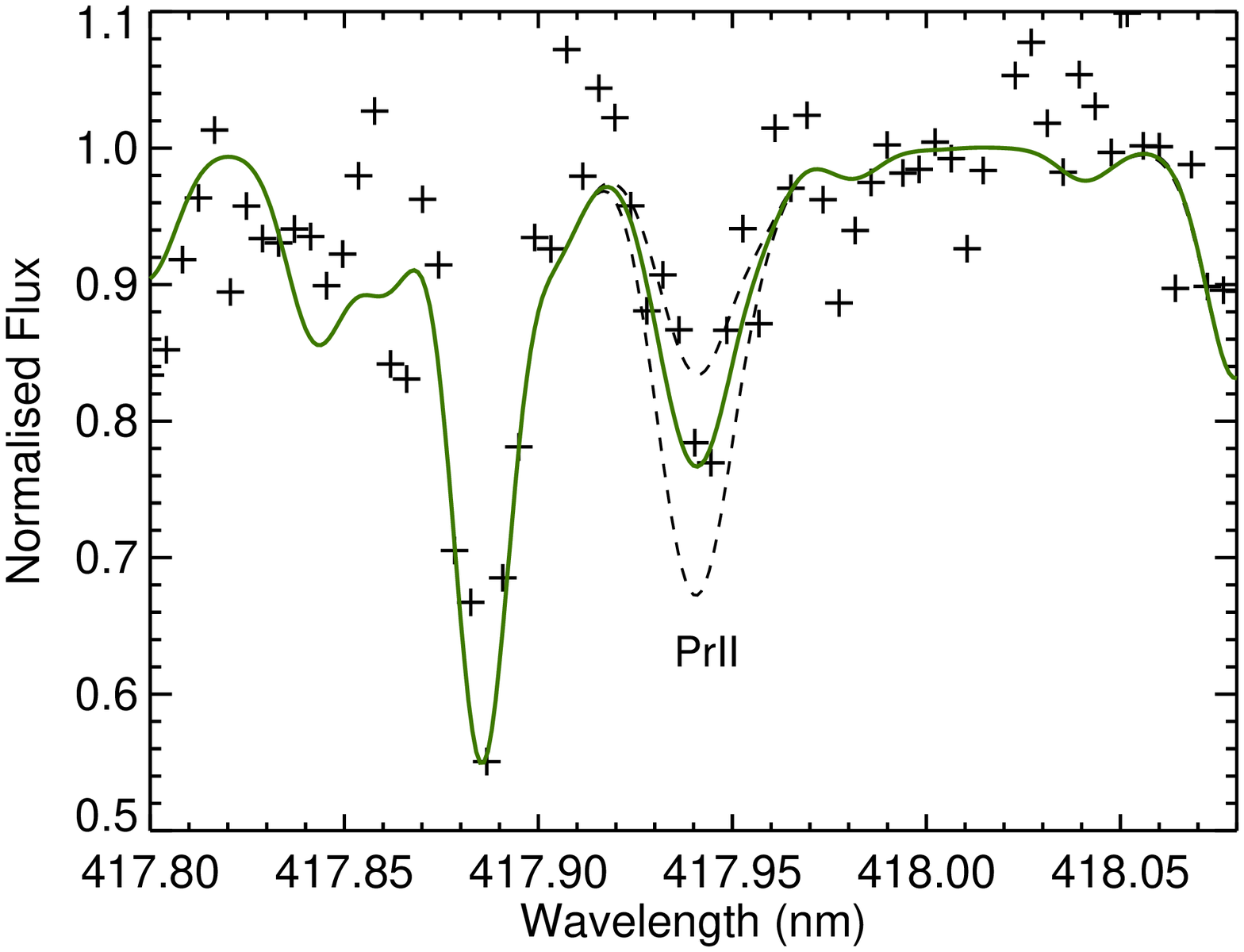}
\includegraphics[scale=0.28]{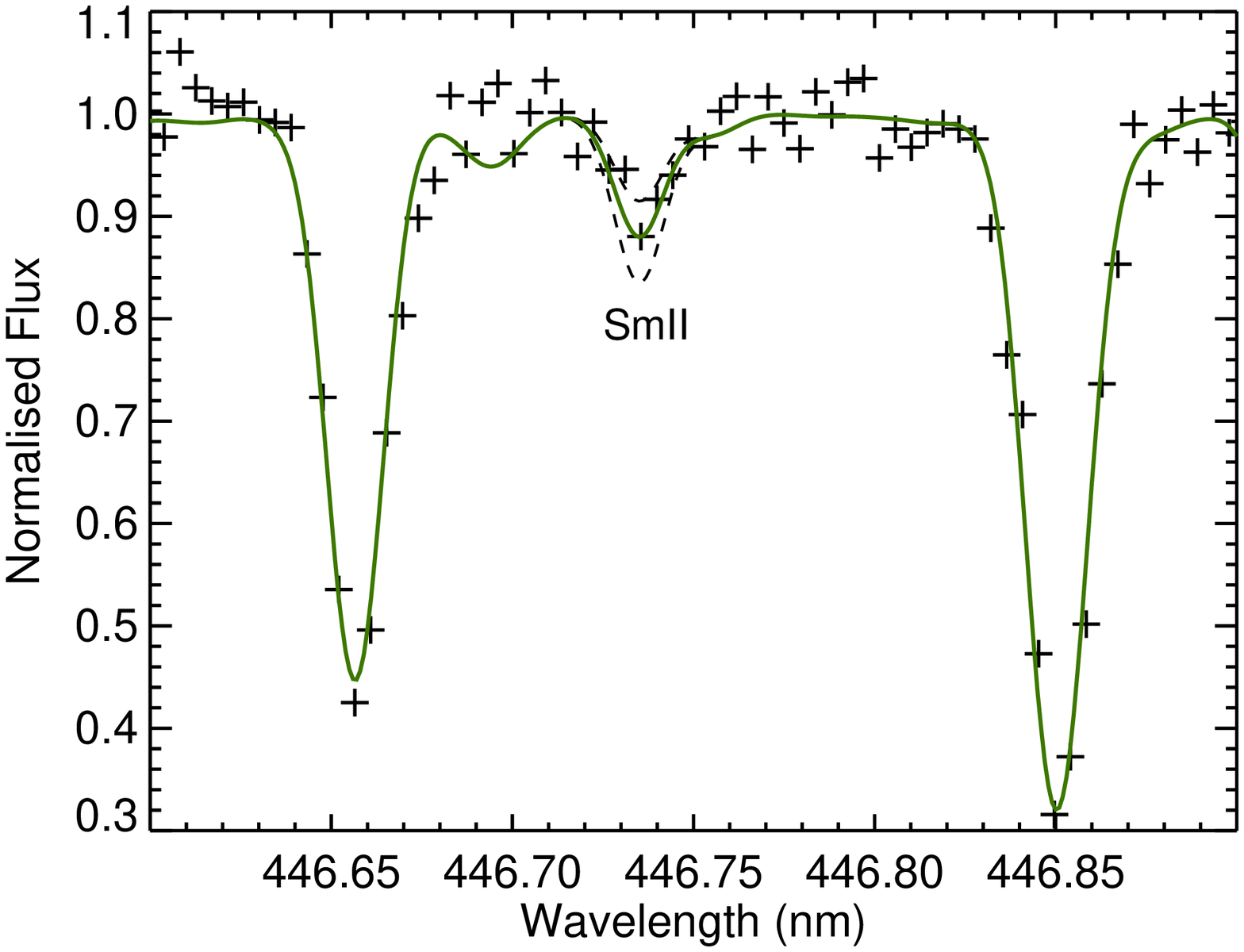}
\includegraphics[scale=0.28]{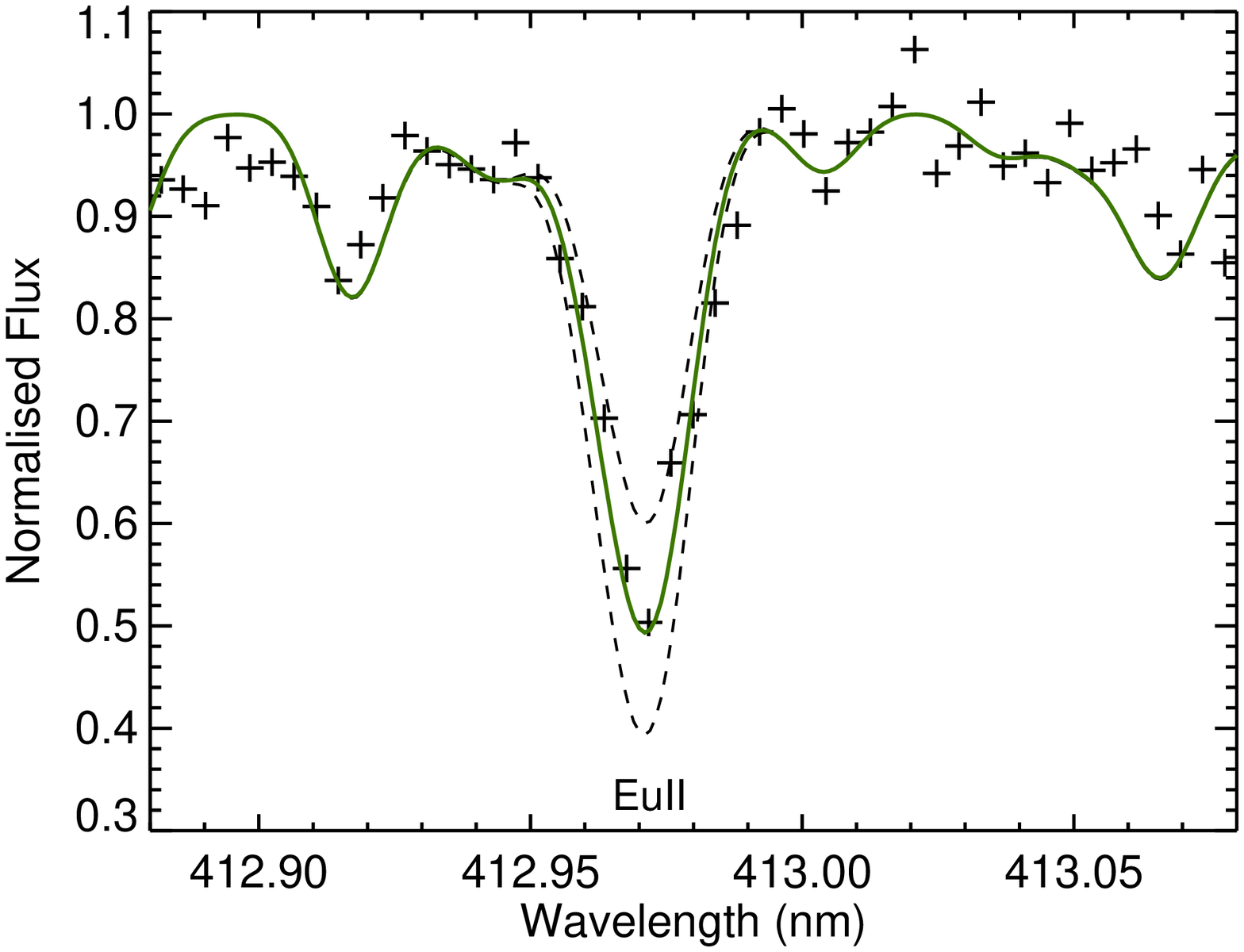}
\end{center}
\caption{Spectrum synthesis of the $r$-process elements Pr (left), Sm (middle) and Eu (right) in DES~J235532. The green solid lines show the spectrum fit and dashed lines show $\mathrm{[X/Fe]} \pm \sigma_{\mathrm{[X/Fe]}}$ \label{figspec}} 
\end{figure*}

The abundances have been derived via spectrum synthesis using the 2011 version of the LTE spectral synthesis program MOOG \citep{sneden1973}, which includes proper treatment of continuum scattering \citep{sobeck2011}. We used an $\alpha$-enhanced $\mathrm{[\alpha/Fe]} = 0.4$ ATLAS9 model atmosphere \citep{castelli2003}. The cool temperature of this star along with its carbon and iron abundances ($\mathrm{[C/Fe]} = -0.25$ and $\mathrm{[Fe/H]} = -2.25$) results in many of the absorption lines being blended, mainly due to the presence of many molecular CH lines in the spectrum, hence an abundance analysis including all elements detected in the spectrum can only be done with spectrum synthesis. Excitation potentials and oscillator strengths for the lines employed are taken from the VALD database \citep{ryabchikova2015}, except for Ba where values from \cite{fuhr2009} were used. For lines exhibiting isotopic shifts and/or hyperfine splitting (Sc, Mn, Co, Cu, Ba, La, Pr, Nd, Sm, Eu and Er) the isotope ratios and hyperfine structure collected by \citet{roederer2014a} were used. Lines used in the abundance analysis are listed in Table \ref{tab:atomic}. Molecular line lists for CH, NH and CN are from \citet{masseron2014} and T. Masseron (private comm.), assuming dissociation energies of 3.47 eV, 3.42 eV and 7.74 eV for the species CH, NH and CN respectively. The C abundance was determined from the CH G-band at 4300~\AA\ and the N abundance from the CN band at 3890~\AA\ assuming an oxygen abundance of $\mathrm{[O/Fe]} = 0.4$. We use solar abundances from \citet{asplund2009}. Figure \ref{figspec} shows portions of the spectrum of DES~J235532 along with syntheses of the most prominent lines of three $r$-process elements: Pr ($\lambda 4179$~\AA), Sm ($\lambda 4467$~\AA), and Eu ($\lambda 4129$~\AA). 

\begin{deluxetable}{lrrrrr}
\tablecaption{Atomic line data.\label{tab:atomic}}
\tablewidth{0pt}
\tabletypesize{\scriptsize}
\tablehead{\colhead{Species} & \colhead{$\lambda$} & \colhead{$\chi$} & \colhead{$\log$ $gf$} \\
\colhead{} &\colhead{(\AA)} & \colhead{(eV)} & \colhead{(dex)} }
\startdata
O~I & 6300.30  & 0.00 & -9.78\\
Na~I & 5889.95 & 0.00 &  0.11\\
Na~I & 5895.92 & 0.00 & -0.19\\
Mg~I & 3829.36 & 2.71 & -0.21\\
Mg~I & 3832.30 & 2.71 &  0.13\\
Mg~I & 3838.29 & 2.72 &  0.40\\
Mg~I & 4167.27 & 4.35 & -0.71\\
Mg~I & 5172.68 & 2.71 & -0.45\\
Mg~I & 5183.60 & 2.72 & -0.24\\
Mg~I & 5528.40 & 4.34 & -0.50\\
Al~I & 3944.00 & 0.00 & -0.64\\
Si~I & 3905.52 & 1.91 & -1.04\\
Si~I & 4102.94 & 1.91 & -3.14\\
\enddata
\tablecomments{
The complete version of Table~\ref{tab:atomic} is available online only.
A short version is shown here to
illustrate its form and content.
}

\end{deluxetable}

\section{Results}
\label{results}

\subsection{Abundance Pattern of DES~J235532}
\label{des_abund}

We have derived abundances or useful upper limits for 30 elements in DES~J235532, including 13 neutron-capture species. Final abundances with uncertainties arising from stellar parameter uncertainties, line information, and continuum placement are listed in Table \ref{tab2}. Uncertainties from stellar parameters were computed by deriving abundances with different atmosphere models, each with one parameter varied by its uncertainty as quoted in \S\ref{analysis}.  These uncertainties were then added in quadrature to the estimated 0.1~dex uncertainty from line information and continuum placement. DES~J235532 exhibits a clear enhancement in Europium
($\mathrm{[Eu/Fe]} = 0.60$) and other neutron-capture elements primarily produced in the $r$-process and can therefore be classified as an $r$-I star. 
 
\begin{deluxetable}{lr}
\tabletypesize{\scriptsize}
\tablecaption{Stellar Parameters and Derived Abundances for DES~J235532 \label{tab2}}
\tablewidth{0pt}
\tablehead{
\colhead{} & \colhead{} }
\startdata
RA & 23:55:32.8  \\
Dec & -59:31:12.0  \\
v$_{{\rm hel}}$ ($\pm 0.3$ km s$^{-1}$) &  $-103.4$\\
\cutinhead{Parameters}
$T_{\rm eff}$ ($\pm$100~K)     & 4720   \\
$\log g$ ($\pm$0.3~dex)       & 1.33   \\
$\mathrm{[Fe/H]}$ ($\pm$0.18~dex)     & $-2.25$  \\ 
$\xi$ ($\pm$0.3~km s$^{-1}$) & 2.0   \\
\cutinhead{Abundances $\mathrm{[X/Fe]}$}
C  &   $-$0.25 (0.25)  \\
N  &   $+$0.10 (0.26)  \\
O  &$<$$+$0.40\nodata  \\
Na &   $+$0.19 (0.30)  \\
Mg &   $+$0.39 (0.29)  \\
Al &   $-$0.86 (0.30)  \\
Si &   $+$0.52 (0.31)  \\
K  &   $+$0.29 (0.19)  \\
Ca &   $+$0.05 (0.20)  \\
Sc &   $+$0.01 (0.29)  \\
Ti &   $+$0.10 (0.28)  \\
Cr &   $-$0.45 (0.19)  \\
Mn &   $-$0.57 (0.25)  \\
Co &   $+$0.13 (0.27)  \\
Ni &   $-$0.09 (0.22)  \\
Cu &$<$$-$0.60\nodata  \\
Zn &   $-$0.02 (0.18)  \\
Sr &   $-$0.39 (0.27)  \\
Y  &   $-$0.44 (0.18)  \\
Zr &   $+$0.05 (0.21)  \\
Ba &   $-$0.16 (0.18)  \\
La &   $+$0.04 (0.20)  \\
Ce &   $-$0.19 (0.17)  \\
Pr &   $+$0.56 (0.23)  \\
Nd &   $+$0.21 (0.23)  \\
Sm &   $+$0.56 (0.18)  \\
Eu &   $+$0.60 (0.19)  \\
Gd &   $+$0.53 (0.20)  \\
Dy &   $+$0.65 (0.21)  \\
Er &   $+$0.61 (0.23)  
\enddata
\end{deluxetable}

Apart from neutron-capture species, the abundances of DES~J235532 are
generally unremarkable. Its metallicity of $\mathrm{[Fe/H]} = -2.25$
places it slightly above the mean metallicity of Tuc~III, and is consistent within the uncertainties with the value calculated by \citet{simon2016} using the Ca triplet method. The $\alpha$ elements in DES~J235532 are
enhanced relative to solar ratios, as expected for its metallicity, with the possible exception of a moderately low Ca abundance ($\mathrm{[Ca/Fe]}$ is consistent with a normal enhancement within the uncertainties). DES~J235532 is not enhanced in carbon.

The nature of Tuc~III was discussed by \citet{simon2016}, who suggested that the system is most likely a dwarf galaxy, although they could not definitively rule out a globular cluster classification. The abundance pattern of the stars in Tuc~III could potentially provide additional insight into this question. However, with only one star we can not determine whether Tuc~III contains the typical light element abundance correlations seen in cluster stars. We note that $r$-process enhanced stars have been found in both clusters and in dwarf galaxies \citep{roederer2011,yong2013}. For the remainder of this paper we assume that Tuc~III is a dwarf galaxy, as favored by the evidence presented by \citet{simon2016}.

\subsection{$r$-process Enhanced Stars in Dwarf Galaxies}
\label{sec:rI_dwarf}

As mentioned in \S\ref{sec:intro}, a number of $r$-process
enhanced stars have been found in other dwarf galaxies, namely in the
classical dwarf spheroidals (dSphs) UMi, Draco, Sculptor, Fornax, and Carina, as well as Ret~II. No $r$-process enhanced stars in ultra-faint dwarfs were known prior to the discovery of Ret~II; \citet{francois2016} found a few stars in Hercules, Canes Venatici~I, and Canes Venatici~II with high Sr and/or Ba abundances, but without measurements of other neutron-capture elements it is not clear whether these stars were enriched by the $r$-process or the $s$-process.

To provide a comparison set for DES~J235532, in this section we compile a sample of all known $r$-process enhanced stars in dwarf galaxies from the literature. Our selection criteria, which will be explored further in \S\ref{sec:s_pollution}, are $\mathrm{[Eu/Fe]} > 0.3$ and $\mathrm{[Eu/Ba]} > 0.4$. In the following, we examine Eu measurements for stars in the individual galaxies. The final selected $r$-I ($0.3 < \mathrm{[Eu/Fe]} < 1.0$) and $r$-II ($\mathrm{[Eu/Fe]} > 1.0$) stars are listed in Tables \ref{tab3} and \ref{tab4}, respectively.\\

\noindent{\it Ursa Minor}

In UMi, 17 giant stars have been analysed with high resolution spectroscopy
\citep{shetrone2001,sadakane2004,cohen2010,kirby2012}. Of the ten stars in the sample of \citet{cohen2010}, five can be labeled as $r$-I stars. Four of the remaining stars have subsolar Eu abundances, while for one only an upper limit is available. All stars in this sample with measured Eu have lower Ba abundance than Eu abundance. One $r$-I star and one $r$-II star were found among the six stars studied by \citet{shetrone2001}. Additionally, two other stars from \citet{shetrone2001} show $\mathrm{[Eu/Fe]} = 0.26 \pm 0.2$ and $0.15 \pm 0.20$, consistent with a solar $\mathrm{[Eu/Fe]}$ ratio or a slight Eu enhancement, and a third has $\mathrm{[Eu/Fe]} = 1.04$ while also having $\mathrm{[Ba/Fe]} = 1.37$. The $\mathrm{[Ba/Fe]}$ ratios range from $-0.39$ to 1.37, but aside from the star with highest Ba abundance, all have higher Eu abundances. Two of the three stars studied by \citet{sadakane2004} overlap with the sample of \citet{shetrone2001} and were already identified as $r$-process enhanced there. For the third star they find $\mathrm{[Ba/Fe]} = -1.28$, but no Eu abundance is given. Finally, \citet{kirby2012} find $\mathrm{[Ba/Fe]} = -0.93$ and an upper limit of $\mathrm{[Eu/Fe]} < 0.23$ for the one UMi star they observed. Hence of the 17 UMi stars for which Eu measurements are available, ten have super-solar $r$-process abundance ratios ($\mathrm{[r/Fe]} > 0$) and seven qualify as $r$-I or $r$-II stars. \\

\noindent{\it Draco}

For the Draco dSph, six giants were analysed by \citet{shetrone2001}, who found two $r$-I stars, and a further eight stars were studied by \citet{cohen2009}, including two additional $r$-I stars, one $r$-II star, and one star with $\mathrm{[Eu/Fe]} = 0.19 \pm 0.19$. Of the six stars analysed in \citet{shetrone2001}, $\mathrm{[Ba/Fe]}$ ratios were derived for five, ranging from $-$1.19 to 0.41, with the two $r$-I stars at $\mathrm{[Ba/Fe]} =$ 0.11 and 0.09. Ba abundances were derived for all eight stars studied by \citet{cohen2009}, with the majority having subsolar Ba abundances. The $r$-II star with $\mathrm{[Ba/Fe]} = 0.26$ shows the highest Ba abundance in the sample. \citet{tsujimoto2015} also analysed three stars in Draco and found one star with a small Eu enhancement of 0.12~dex and $\mathrm{[Ba/Fe]} = -0.69$. Therefore, seven of the 17 stars in Draco with high resolution spectra in the literature are enhanced in $r$-process elements, with five qualifying as $r$-I or $r$-II stars. For the remaining ten objects, two have upper limits of $\mathrm{[Eu/Fe]} < 0.12$ and $0.41$, while eight stars have no Eu abundance measured.\\ 

\noindent{\it Sculptor}

Seven papers report Eu abundances or upper limits for a total of 21 stars in Sculptor \citep{shetrone2003,geisler2005,tafelmeyer2010,kirby2012,simon2015,jablonka2015,skuladottir2015}. Eight of these stars show enhancement in $r$-process elements, including one $r$-I star from the sample of \citet{kirby2012}. The other seven stars with Eu enhancement (four from \citealt{shetrone2003} and three from \citealt{geisler2005}) are also enhanced in the $s$-process element Ba. Of the remaining 13 stars, three have subsolar Eu detection or upper limits \citep{geisler2005,jablonka2015}, nine have upper limits above solar that do not provide significant constraints
\citep{tafelmeyer2010,simon2015,jablonka2015}, and one is identified as a
CEMP-no star\footnote{Carbon enhanced metal-poor star with no enhancement in neutron-capture elements \citep{beers2005}.} \citep{skuladottir2015}. In summary only one of the 21 stars in Sculptor with Eu measurements qualifies as an $r$-I star and none are $r$-II stars.\\

\noindent{\it Fornax}

The dSph Fornax is well studied, with neutron-capture element abundances published for 147 stars \citep{shetrone2003,tafelmeyer2010,letarte2010,lemasle2014}. The
majority of these are enhanced not only in Eu but also in $s$-process elements such as Ba and La, such that only six qualify as $r$-I stars; three from the sample of \citet{lemasle2014} and three from the sample of \citet{letarte2010}. Furthermore one star from \citet{lemasle2014} and three stars from \citet{letarte2010} qualify as $r$-II stars, resulting in a total of 10 $r$-I and $r$-II stars discovered in this galaxy.\\

\noindent{\it Carina}

\citet{shetrone2003} and \citet{venn2012} have published Eu abundances or upper limits for five and nine stars, respectively, in the Carina dSph. Here two stars qualify as $r$-I stars, one from each sample, while the rest either show a similar enhancement in $s$-process elements or only have an upper limit on the Eu abundance.\\ 

\noindent{\it Reticulum II}

Nine stars in Ret~II have been analysed with high resolution spectroscopy, and seven of these show large enhancements in $r$-process elements \citep{ji2016,ji2016b,roederer16}. One of the $r$-process enhanced stars in Ret~II also has a high Ba abundance ($\mathrm{[Eu/Ba]} = 0.3$) and does not meet our selection criteria. The remaining six stars are listed in Table~\ref{tab4}.

A number of $r$-process enhanced stars in the halo also show enhancements in carbon \citep[e.g., CS~22892$-$052;][]{sneden2003} and overall carbon enhancement is the most common abundance peculiarity in metal-poor stars. All of the stars listed in Table \ref{tab2} and \ref{tab3} are evolved giants and thus some change in their original carbon abundances is expected from stellar evolution. We have performed carbon corrections using the tool provided by \citet{placco2014} but even though corrections are substantial ($\sim 0.6-0.7$~dex), none of the $r$-I or $r$-II stars in UMi, Draco, or Sculptor qualify as being carbon-enhanced metal-poor stars (CEMP; $\mathrm{[C/Fe]} \geq 0.7$; \citealt{aoki2007}). For the six Fornax and two Carina stars, no C abundances are available. \citet{ji2016b} performed the same exercise for the $r$-process enhanced stars in Ret~II and found three of them to be carbon enhanced. 

\begin{deluxetable*}{lllll}
\tabletypesize{\scriptsize}
\tablecaption{$\mathrm{[Fe/H]}$, $\mathrm{[Eu/Fe]}$ and $\mathrm{[Eu/Ba]}$ ratios for $r$-I stars identified in dwarf galaxies.\label{tab3}}
\tablewidth{0pt}
\tablehead{
\colhead{Star ID} &\colhead{$\mathrm{[Fe/H]}$} & \colhead{$\mathrm{[Eu/Fe]}$} &\colhead{$\mathrm{[Eu/Ba]}$} & \colhead{Ref}}
\cutinhead{Tuc~III}
 DES~J235532 & -2.25 & 0.60 &0.76 & This paper \\
\cutinhead{UMi}
N37    & -1.55 & 0.87 &0.68 & \citet{cohen2010} \\
J12    & -1.76 & 0.83 &0.56 & \citet{cohen2010}\\
297    & -1.68 & 0.74 &0.59 & \citet{shetrone2001}\\
27940  & -1.91 & 0.61 &0.66 & \citet{cohen2010}\\
COS233 & -2.15 & 0.52 &0.70 & \citet{cohen2010} \\
36886  & -2.43 & 0.44 &0.79 & \citet{cohen2010} \\
\cutinhead{Draco}
 3150 & -1.89 & 0.84 &0.70 & \citet{cohen2009} \\
 11   & -1.72 & 0.55 &0.48 & \citet{shetrone2001}\\
 343  & -1.86 & 0.51 &0.45 & \citet{shetrone2001}\\
 3053 & -1.73 & 0.39 &0.54 & \citet{cohen2009} \\
\cutinhead{Sculptor}
1019417 & -2.45 & 0.39 &0.83 & \citet{kirby2012}\\
\cutinhead{Fornax}
mem0612  & -1.68 & 0.93 &1.22 & \citet{lemasle2014} \\
rgb 0522 & -1.87 & 0.76 &1.11 & \citet{lemasle2014}\\
mem0747  & -1.39 & 0.68 &0.67 & \citet{lemasle2014}\\
BL196	 & -1.06 & 0.59 &0.42 & \citet{letarte2010}\\
BL190	 & -0.79 & 0.54 &0.48 & \citet{letarte2010}\\
BL228    & -0.88 & 0.41 &0.46 & \citet{letarte2010}\\
\cutinhead{Carina}
Car 10 & -1.94 & 0.80 & 0.55 & \citet{shetrone2003}\\
1013   & -1.30 & 0.33 & 0.48 & \citet{venn2012}
\enddata
\end{deluxetable*} 

\begin{deluxetable*}{lllll}
\tablecaption{$\mathrm{[Fe/H]}$, $\mathrm{[Eu/Fe]}$ and $\mathrm{[Eu/Ba]}$ ratios for $r$-II stars identified in dwarf galaxies.\label{tab4}}

\tablewidth{0pt}
\tablehead{
\colhead{Star ID} &\colhead{$\mathrm{[Fe/H]}$} & \colhead{$\mathrm{[Eu/Fe]}$} &\colhead{$\mathrm{[Eu/Ba]}$} & \colhead{Ref}}
\cutinhead{UMi}
199 & -1.45 & 1.49 & 0.72 & \citet{shetrone2001}\\
\cutinhead{Draco}
21456 & -2.34 & 1.05 & 0.79 & \citet{cohen2009}\\
\cutinhead{Fornax}
BL147  & -1.38 & 1.72 & 0.43 & \citet{letarte2010}\\
BL204  & -1.00 & 1.26 & 0.80 & \citet{letarte2010} \\
BL311  & -0.78 & 1.24 & 0.61 & \citet{letarte2010} \\
mem0704 & -2.55 & 1.10 & 0.70 & \citet{lemasle2014}\\
\cutinhead{Reticulum II}
DES J033454-540558 & -2.77 & 2.11 & 0.71 &\citet{ji2016b}\\
DES J033447-540525 & -2.91 & 1.86 & 0.78 &\citet{ji2016b}\\
DES J033457-540531 & -2.08 & 1.76 & 0.40 &\citet{ji2016b}\\
DES J033607-540235 & -2.97 & 1.74 & 0.83 &\citet{ji2016b}\\
DES J033523-540407 & -3.01 & 1.68 & 0.89 &\citet{ji2016b}\\
DES J033548-540349 & -2.19 & 0.95 & 0.59 &\citet{ji2016b}
\enddata
\end{deluxetable*}

\subsection{Pollution From $s$-process Nucleosynthesis}
\label{sec:s_pollution}

Europium is mainly produced in the $r$-process, while Barium is mainly produced in the $s$-process in asymptotic giant branch (AGB) stars, so the ratio of the two is a diagnostic of the onset of the $s$-process, seen as a decline in the $\mathrm{[Eu/Ba]}$ ratio with increasing metallicity. \citet{simmerer2004} found an $r$-process ratio of $\mathrm{[Eu/Ba]} = 0.70$ for the Solar System. The two classic halo $r$-II stars CS~31082$-$001 \citep{siqueira2013} and CS~22892$-$052 \citep{roederer2014a}, which are often used as templates of the $r$-II abundance pattern, have slightly lower abundance ratios of $\mathrm{[Eu/Ba]}= 0.53$ and $\mathrm{[Eu/Ba]} = 0.41$, respectively.
Based on these measurements we included a limit on the $\mathrm{[Eu/Ba]}$ ratio of $\mathrm{[Eu/Ba]} > 0.4$ to ensure that the heavy element abundance patterns of the neutron-capture-enhanced stars selected in \S\ref{sec:rI_dwarf} are dominated by the $r$-process contribution and have not been severely polluted by $s$-process material. $\mathrm{[Eu/Ba]}$ ratios for all of the selected $r$-I and $r$-II stars are listed in Tables~\ref{tab3} and \ref{tab4}. For DES~J235532 we find $\mathrm{[Eu/Ba]} = 0.76$, similar to what is found for the $r$-I and $r$-II stars in the other dwarfs.

\citet{kirby2012} detected no decline in the $\mathrm{[Eu/Ba]}$ ratio with increasing metallicity in UMi after combining the samples of \citet{shetrone2001,sadakane2004} and \citet{cohen2010}. The authors suggest that the $\mathrm{[Eu/Ba]}$ ratio is constant because star formation in UMi had stopped by the time that its AGB stars began to pollute the ISM, so the $s$-process elements created there were never incorporated into any new stars. However, we note that (as mentioned above) one UMi star with $\mathrm{[Fe/H]} = -2.17$, from the sample of \citet{shetrone2003}, does show a large enhancement in Ba in addition to its Eu enhancement. Also, the six $r$-I stars in UMi listed in Table~\ref{tab3} do show a small decline in $\mathrm{[Eu/Ba]}$ with increasing metallicity.

In Fornax and Sculptor, on the other hand, the $\mathrm{[Eu/Ba]}$ ratio starts to decline at $\mathrm{[Fe/H]} = -1.5$ \citep{lemasle2014} and $\mathrm{[Fe/H]} \sim -1.8$ \citep{kirby2012}, respectively, and \citet{cohen2009} detect a decline at similar metallicity in Draco. The majority of our selected $r$-I stars have metallicities similar to or lower than these values, supporting the idea that the neutron-capture elements in these stars were primarily produced by the $r$-process.

\subsection{$r$-I Abundance Pattern in Dwarf Galaxies}

To constrain the origin of metal-poor $r$-process enhanced stars, in this subsection we compare their abundance patterns to those of ordinary metal-poor halo stars and to the $r$-process abundance pattern found in the Sun.

\subsubsection{Light elements}

\begin{figure}
\center
\includegraphics[scale=0.45]{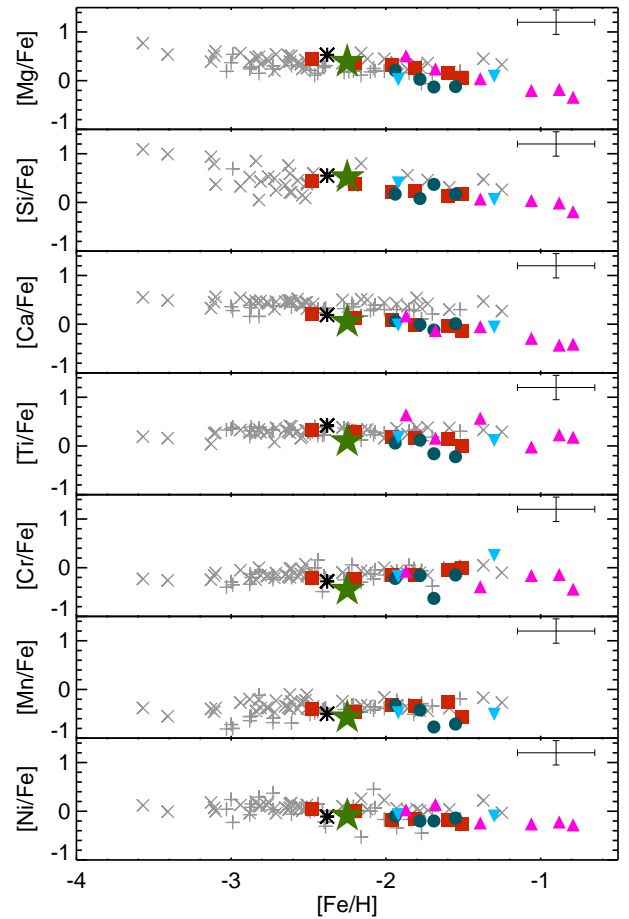}
\caption{Mg, Si, Ca, Ti, Cr, Mn and Ni abundances for $r$-I stars in dwarf galaxies and in the halo. A typical error bar is displayed in the upper right corner of each panel. The plotted sample includes Tuc~III (green star), UMi (red squares), Draco (dark blue circles), Sculptor (black  asterisks), Carina (light blue upside-down triangles), Fornax (magenta triangles), and halo stars (grey plusses from \citealt{barklem2005} and crosses from
  \citealt{roederer2014a}). References for literature dwarf galaxy data are given in Table \ref{tab2}. \label{fig2}}  
\end{figure}

\citet{siqueira2014} and \citet{roederer2014b} explored abundances for the light elements from Mg to Ni in $r$-I and $r$-II stars (respectively) and found no significant difference between the abundances in $r$-process enhanced stars and normal halo stars. 

Figure \ref{fig2} shows the abundances for the $\alpha$ and iron peak elements in DES~J235532 compared with $r$-I stars from the halo \citep{roederer2014a,barklem2005} and $r$-I stars from UMi, Draco, Sculptor, Fornax and Carina. All abundances have been converted to the \citet{asplund2009} solar abundance scale. For the majority of the $\alpha$ and iron peak elements, similar abundances are found in the halo $r$-I and dwarf galaxy $r$-I stars. The possible exception is Ca, where as noted in \S\ref{des_abund}, DES~J235532 has a slightly lower abundance than most halo stars at similar metallicity. The other dwarf galaxy $r$-I stars also have on average $\sim$0.2~dex lower $\mathrm{[Ca/Fe]}$ abundances than the halo $r$-I stars. A similar trend may be seen at higher metallicities for Ti with lower significance. It should be noted, though, that the $r$-process enhanced stars in the halo cover a wider metallicity range than those in dwarf galaxies, with the majority having $\mathrm{[Fe/H]} < -2.5$, mainly due to selection effects in halo surveys, while the dwarf galaxy stars are all at $\mathrm{[Fe/H]} > -2.5$. Thus, it would be interesting to examine a larger sample of higher-metallicity halo $r$-I stars to see if their Ca and Ti abundances are indeed distinct from the dwarf galaxy $r$-I stars.

\subsubsection{Neutron-capture elements}

It was recognized from early on that for the highly $r$-process enhanced $r$-II stars in the halo the heavy neutron-capture element ($Z\geq56$) abundance ratios match those of the Solar System \citep{mcwilliam1995,barklem2005}. Recently this result was found to extend to dwarf galaxy stars when \citet{ji2016} and \citet{roederer16} found an equally good match for the $r$-II stars discovered in Ret~II. Furthermore, \citet{siqueira2014} compared the abundance pattern from Sr to Th of halo $r$-I stars to that of the halo $r$-II star CS~31082$-$001, finding good agreement for the heavy elements (Ba to Th), while the light neutron-capture elements Sr, Y, and Zr were more enhanced in the $r$-I stars relative to CS~31082$-$001. Thus the heavy neutron-capture element abundance pattern seen in all $r$-process enhanced stars in the halo and Ret~II matches that of the Solar System.
   
\begin{figure}
\includegraphics[scale=0.59]{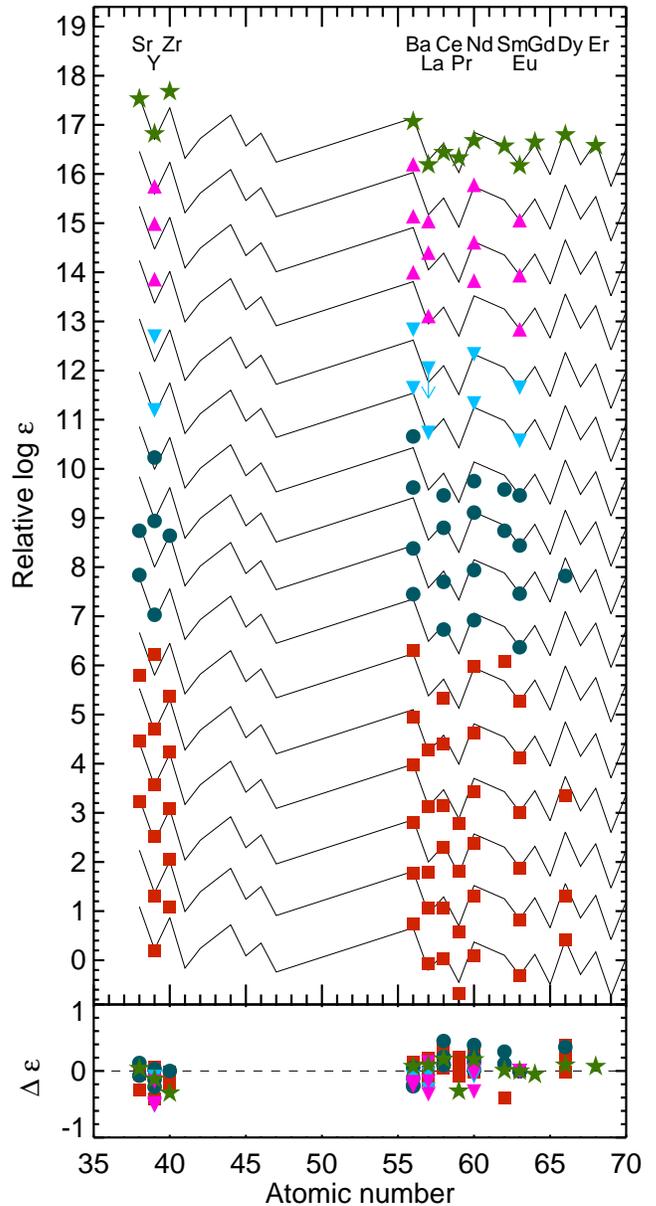}
\caption{Scaled neutron-capture element abundances for dwarf galaxy $r$-I stars compared to the abundance pattern of CS~22892$-$052 (solid line). The bottom panel shows the individual stellar abundance offsets with respect to CS~22892$-$052 after normalizing to the Eu abundance. The green stars show the derived abundances for DES~J235532 in Tuc~III from this paper. Stars in other dwarf galaxies from the literature are plotted as magenta triangles (Fornax), light blue upside-down triangles (Carina), dark blue circles (Draco), and red squares (UMi). References for the literature data are given in Table \ref{tab2}. The height of the plotted points is representative of the abundance uncertainties and the arrow for the first Carina star indicates an upper limit.\label{fig1}} 
\end{figure}

In Figure~\ref{fig1} we compare the neutron-capture element absolute abundance pattern for the $r$-I stars detected in the aforementioned dwarf galaxies, along with DES~J235532, with the abundance pattern of the
classical $r$-II star CS~22892$-$052 taken from \citet{sneden2003}, normalized to the Eu abundance. Also shown is a difference plot with the stellar abundance differences between the individual dwarf galaxy stars and CS~22892$-$052. The abundance uncertainties are comparable to the height of the plotted symbols. For the Sculptor $r$-I stars only upper limits were reported for the heavy neutron-capture elements besides Eu and Ba, making it difficult to determine whether the abundance pattern follows that of CS~22892$-$052, hence these stars have not been included in Figure~\ref{fig1}. The same is also the case for three of the six $r$-I stars in Fornax, therefore only BL190, BL196, and BL228 have been included. For the remainder, abundances have been reported for four or more elements. As can be seen there is good overall agreement for the heavy elements between the abundance pattern of CS~22892$-$052 and the abundance patterns of the $r$-process enhanced stars found in dwarf galaxies, with an average abundance difference of only 0.16~dex for the elements from Ba to Er. In particular, the result of \citet{siqueira2014} for the $r$-I stars in the halo --- that the abundance pattern follows the same $r$-process pattern as the $r$-II stars --- is also valid for $r$-I stars in dwarf galaxies.

\subsection{Source of $r$-Process Enrichment in Dwarf Galaxies}

\citet{tsujimoto2014} examined literature data for Eu abundances for stars in a number of dwarf galaxies (Carina, Draco, Sculptor, Fornax, Sagittarius [Sgr], Ursa Major II, Hercules, and Coma Berenices) and found different behavior for this element between the various galaxies. Some systems exhibit
a constant Eu level of $\mathrm{[Eu/H]}\sim -1.3$ over the metallicity range $(-2 \la \mathrm{[Fe/H]} \la -1)$, while others show a steady increase in Eu with metallicity. The authors argue that this result can be explained if the $r$-process enrichment is due to NSMs, where the galaxies showing a plateau have on average experienced $< 0.1$ NSM and were enriched by gas from a more massive neighbor, while galaxies with an increasing trend in Eu with Fe have hosted multiple NSMs. Carina and Sculptor are examples of the galaxies with apparent Eu plateaus. Larger galaxies such as Fornax and Sgr are examples of the latter (no Eu plateau with increasing metallicity), probably because their stellar masses are large enough that multiple NSMs are likely to have occurred.  Draco, which has a smaller stellar mass, contains low metallicity stars with very low Eu abundance, such as Draco 119 with $\mathrm{[Eu/H]} < -2.55$ \citep{fulbright2004}. \citet{tsujimoto2015} derived Eu abundances for two additional stars in Draco. In combination with the literature sample, these data suggest that stars with $\mathrm{[Fe/H]} < -2.2$ have no Eu enrichment, followed by a sudden increase of 0.7~dex in the Eu abundance for stars with $\mathrm{[Fe/H]} > -2.2$. This pattern led \citet{tsujimoto2015} to reject the theory of pre-enrichment for Draco and suggest that a NSM occurred in the galaxy once its metallicity reached $\mathrm{[Fe/H]} \sim -2.2$, although that hypothesis may not be consistent with Draco's gas accretion history \citep{ji2016}.

\begin{figure*}
\center
\includegraphics[scale=0.6]{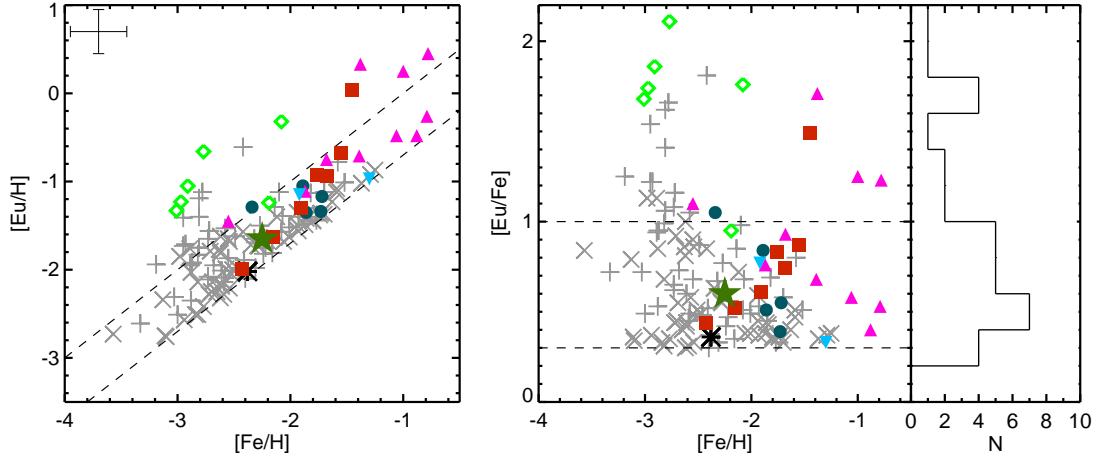}
\caption{$\mathrm{[Eu/H]}$ abundances (left) and $\mathrm{[Eu/Fe]}$ abundances and distribution (right) for $r$-I and $r$-II stars in dwarf galaxies and in the halo. The $\mathrm{[Eu/Fe]}$ histogram in the right panel only includes stars in dwarf galaxies. The plotted sample includes Tuc~III (green star), UMi (red squares), Draco (dark blue circles), Sculptor (black  asterisks), Carina (light blue upside-down triangles), Fornax (magenta triangles), and Ret~II (light green diamonds; \citealt{ji2016b}). The halo stars shown here are taken from \citet[][grey plusses]{barklem2005} and \citet[][grey crosses]{roederer2014a}. References for literature data are given in Tables \ref{tab2} and \ref{tab3}. Dashed lines show limits for $r$-I and $r$-II stars. \label{fig4}}   
\end{figure*}

In Figure \ref{fig4} we plot the $\mathrm{[Eu/H]}$ ratios (left) and the $\mathrm{[Eu/Fe]}$ ratios (right) as a function of metallicity for the $r$-I and $r$-II stars compiled in \S\ref{sec:rI_dwarf} along with the $r$-II stars found in Ret~II and the $r$-I star presented in this paper; the right plot also shows the distribution of $\mathrm{[Eu/Fe]}$ ratios. Most strikingly, the stars in the low mass galaxy Ret~II (green diamonds) show a clear increase in Eu with Fe. A similar trend is seen for the stars in UMi (red squares), with a stellar mass two orders of magnitude higher (and a comparable mass to Draco). These two galaxies also follow the general trend of $\mathrm{[Eu/H]}$ for $r$-process enhanced stars in the halo. DES~J235532 (green star) has a $\mathrm{[Eu/H]}$ ratio placing it in the middle of the halo stars and at the lower end of the dwarf galaxy stars. With the expanded sample of stars from Tables \ref{tab2} and \ref{tab3} the low-metallicity plateau at $\mathrm{[Eu/H]} \sim -1.3$ seen by \citet{tsujimoto2014} for the low luminosity systems is no longer apparent.

Based on the increase seen in $\mathrm{[Eu/H]}$ with $\mathrm{[Fe/H]}$ in Fornax and Sgr, \citet{tsujimoto2014} conclude that more than one NSM must have occurred in these galaxies. However, the data for Ret~II indicate that the same signature can be produced with inhomogeneous mixing and a single NSM \citep{ji2016}. Also, Fornax and Sgr have both been found to be enriched in $s$-process elements, potentially providing an alternative source for the additional Eu observed in their higher metallicity stars \citep{mcwilliam2013,letarte2010}, hence the Eu abundances is not necessarily a clean tracer of the $r$-process in these galaxies.

\section{Discussion}
\label{discussion}

The overall good agreement between the heavy element abundance patterns of
DES~J235532, of $r$-I stars from other dwarf galaxies, and that of CS~22892$-$052, along with the similarly good agreement found when comparing the $r$-II stars in Ret II to CS~22892$-$052 \citep{ji2016b,roederer16}, suggests a common astrophysical source for the $r$-process elements in $r$-I and $r$-II stars in both the halo and dwarf galaxies. As argued by \citet{ji2016}, this source must be rare and each event must produce large quantities of $r$-process material ($\sim10^{-4.5}$~M$_{\odot}$ of Eu), which is best explained by a NSM or magneto-rotationally driven SN.

The detection of an $r$-I star in the ultra-faint dwarf galaxy Tuc~III raises several questions. If $r$-process nucleosynthesis events are rare, how did a galaxy as small as Tuc~III become enhanced in such elements? And why does it have an $r$-I star with moderate $r$-process enhancement, in contrast to Ret~II, whose $r$-process enhanced stars are all extremely-enhanced $r$-II stars?

We explore two possible scenarios for the $r$-process enrichment of Tuc~III.  \citet{tsujimoto2014} pointed out that because of the high ejecta velocities from neutron star mergers, the $r$-process material they synthesize can spread into large volumes with stopping lengths that exceed the radii of dwarf galaxies. Furthermore, \citet{komiya2016} show that a significant fraction of the NSM ejecta can escape the proto-galaxies entirely, polluting the surrounding intergalactic medium (IGM) and other nearby proto-galaxies.  If Tuc~III formed near other systems, perhaps it was polluted by ejecta from a NSM in a larger nearby galaxy. On the other hand, if a NSM occurred in Tuc~III itself, depending on its gas distribution at the time and the geometry of the explosion, much of the resulting $r$-process debris could have escaped from the system, producing a lower level of enrichment than in Ret~II. Future analysis of a larger sample of stars in Tuc~III will reveal whether the galaxy has an overall enhancement of $r$-process elements seen in many stars, or if DES~J235532 is an outlier.

Alternatively, we note that the $r$-I stars drawn from the literature have mostly been identified in more luminous dwarf galaxies. These systems likely had larger gas reservoirs at early times into which the ejecta from an $r$-process event could be diluted. The observed spread of $r$-process abundances within these galaxies, including both $r$-I and $r$-II stars, may have resulted from incomplete or inhomogeneous mixing. With a stellar mass of M$_* = 800$ M$_\odot$ \citep{drlica-wagner2015}, Tuc~III is the lowest-luminosity dwarf found to contain an $r$-I star. However, there is evidence that Tuc~III was more massive in the past. \citet{drlica-wagner2015} discovered apparent tidal tails on either side of the galaxy (see their Fig. 8). The integrated luminosity of the tails is comparable to the current luminosity of Tuc~III itself, suggesting that at least half of the stellar mass has been lost to tidal stripping.  With current data it is difficult to place an upper limit on the amount of mass that could have been stripped. According to the tidal evolution models of \citet{penarrubia2008}, the current luminosity and velocity dispersion of Tuc~III are consistent with the stripping of a $\sim10^{6}$~L$_{\odot}$ progenitor similar to the classical dSphs \citep{simon2016}, but that would imply the loss of 99.9\%\ of its initial stellar mass. Deeper imaging of the tails of Tuc~III and a proper motion measurement to determine its orbit will be necessary to better constrain its mass loss. Still, for now we can at least conclude that it is plausible that Tuc~III was once a more luminous galaxy than Ret~II (M$_*$ $= 2.6\times10^3$~M$_\odot$; \citealt{bechtol2015}).  We also cannot rule out the possibility that Tuc~III had a luminosity as large as the other dwarfs with $r$-I stars before it ventured too close to the Milky Way.

If abundance measurements for a larger sample of Tuc~III stars show a uniform level of $r$-process enrichment across the galaxy, that would point toward IGM pollution as the source of its $r$-process material. An internal spread in $r$-process abundances would most likely indicate that $r$-process elements were produced in Tuc~III. If Tuc~III contains a range of $r$-I and $r$-II stars similar to UMi and Draco, that may suggest that it was a much larger dwarf galaxy at earlier times before it was tidally stripped by the Milky Way.  On the other hand, if most other Tuc~III stars are not $r$-process enhanced, then much of the material responsible for enriching DES~J235532 was likely ejected from the galaxy. 

The astrophysical source of the $r$-process elements has been narrowed down, but is not yet conclusively identified. The current most promising sites are NSMs and jet-SNe, events that are similarly rare and that are equally able to reproduce the heavy neutron-capture element abundance pattern seen in $r$-process enhanced stars \citep{wanajo2014,winteler2012}. \citet{tsujimoto2014} and \citet{tsujimoto2015} argue that NSMs can produce the Eu abundances seen in dwarf galaxies, with a plateau seen in dwarfs that hosted no more than one NSM event and increasing Eu with Fe in the dwarfs where multiple NSMs have happened. However, an increasing trend is seen in both UMi and Ret~II, which according to the \cite{tsujimoto2015} model should show a plateau in their $\mathrm{[Eu/H]}$ abundances. Furthermore, some of the dwarfs showing an increasing trend in Eu with metallicity also show evidence of enrichment by the $s$-process, and thus the increase in Eu with metallicity in those systems may not be entirely due to $r$-process enrichment. With only one star in Tuc~III we cannot yet say whether the Eu abundances in this galaxy increase with metallicity, sit on a plateau, or behave in a different manner entirely.  However, given its metallicity the $\mathrm{[Eu/H]}$ ratio of DES~J235532 is consistent with those of other $r$-process enhanced stars in dwarf galaxies and in the halo.

The $r$-process enhanced stars from the individual dwarf galaxies each create a sequence in the left panel of Figure \ref{fig4} (Fornax appears to contain two sequences). For the dwarf galaxies containing $r$-I stars these sequences fall on top of the area populated by the halo $r$-I stars, whereas the $r$-II stars in Ret~II are seen at higher $\mathrm{[Eu/H]}$ ratios. This results in the $\mathrm{[Eu/Fe]}$ distribution seen in the right panel where the top peak is dominated by the Ret~II stars. A more gradual transition is seen for the $r$-process enhanced stars in the halo. From the abundance cuts used to select $r$-I ($0.3 < \mathrm{[Eu/Fe]} < 1$, $\mathrm{[Eu/Ba]} > 0.4$) and $r$-II ($\mathrm{[Eu/Fe]} > 1$,  $\mathrm{[Eu/Ba]} > 0.4$) stars one would expect, with a large enough sample, to see a gradual transition between $r$-I and $r$-II stars at different metallicities, as is the case for the halo stars. For the $r$-stars in the dwarf galaxies the location of the $\mathrm{[Eu/H]}$ sequence may be a result of the level of dilution of the $r$-process ejecta. For example in the low luminosity dwarf Ret~II only a small amount of gas was available to dilute the $r$-process ejecta, whereas the higher luminosity system UMi had a larger gas reservoir within which the $r$-process ejecta were mixed, creating similar sequences in $\mathrm{[Eu/H]}$, but with an offset. Thus in principle with a large sample of $r$-process enhanced dwarf galaxies of varying luminosity the gap between the Ret~II stars and the $r$-process enhanced stars from other dwarf galaxies would be filled in. 
 
The current sample of $r$-I and $r$-II stars in dwarf galaxies to our
knowledge comprises 20 $r$-I stars in five different systems and 12 $r$-II stars in four different systems, with the majority of the $r$-II stars located in Ret~II. This ratio of $r$-I stars to $r$-II stars is similar to what is found in the halo, where at present $\sim 45$ $r$-I and $\sim 25$ $r$-II stars have been identified. The $r$-I stars discovered in dwarf galaxies so far seem to cover the more metal-rich end of the metallicity distribution of halo $r$-I stars. Similarly, the $r$-II stars identified in dwarf galaxies that contain $r$-I stars as well have metallicities of $\mathrm{[Fe/H]}\sim -2$. One $r$-II star of similar metallicity ($\mathrm{[Fe/H]} = -1.7$) has been detected in the MW bulge \citep{johnson2013}, but no $r$-II stars at such high metallicities have yet been detected in the MW halo, where the known $r$-II stars are all at $\mathrm{[Fe/H]} \sim -3$ \citep{barklem2005}.

\section{Conclusions}
\label{conclusion}

In this paper we have presented a full abundance analysis of the star DES~J235532.66$-$593114.9 in the ultra-faint dwarf galaxy Tucana~III. We find that DES~J235532 is enhanced in $r$-process elements and classify it as an $r$-I star. This is the first $r$-I star to be identified in a dwarf galaxy of such low luminosity. After compiling a complete literature sample of dwarf galaxy $r$-I stars, we show that the heavy neutron-capture element abundance pattern of this star and $r$-I stars in other dwarf galaxies matches that of the classical $r$-II star CS~22892$-$052. The same match has been detected for $r$-I stars in the halo; it is therefore possible that the $r$-I stars currently located in the Milky Way halo were born in dwarf galaxies covering a range in luminosity such as Tuc~III and Ursa~Minor. 

When comparing the dwarf $r$-I stars to halo $r$-I stars, we find that the majority of the halo stars are more metal-poor and Ca-rich than the $r$-I stars identified in dwarf galaxies. However, the sample size of known $r$-process enhanced stars in both the halo and in dwarf galaxies is still limited and many of the halo studies have been focused on the more metal-poor ($\mathrm{[Fe/H]} < -3.0$) stars. Hence larger samples may reveal lower metallicity $r$-process enhanced stars in dwarfs and higher metallicity
$r$-process enhanced stars in the halo. 

Because of the very low luminosity of Tuc~III it is surprising that the system contains any $r$-process-enhanced stars. Among other faint dwarfs only Reticulum~II hosts such stars, while the other dwarf galaxies with known $r$-I stars all have stellar masses that are orders of magnitude larger. Both internal and external production of the $r$-process material in Tuc~III are possible, but with abundance information for only a single star we cannot distinguish between them. High-resolution spectroscopy of a small number of additional stars in Tuc~III, covering both the tidal tails and the core, will reveal whether, like Ret~II, Tuc~III was the location of a neutron star merger at early times, or if it was polluted by an outside $r$-process event. 

Finally, we examine the $\mathrm{[Eu/H]}$ abundances for the $r$-process enhanced stars in dwarf galaxies and in the halo. It has been suggested that enrichment from NSMs should result in a plateau in the $\mathrm{[Eu/H]}$ abundances for stars in dwarf galaxies. While small samples may result in the appearance of such plateaus in individual objects, the combined sample from many dwarfs is consistent with a steady increase in $\mathrm{[Eu/H]}$ with metallicity.  Parallel sequences in $\mathrm{[Eu/H]}$ for various galaxies, most notably Ret~II and UMi, may correspond to the different levels of dilution of the $r$-process ejecta in each dwarf.  

The discovery of $r$-process enhanced stars in ultra-faint dwarf galaxies  presents us with an environmental constraint on the event(s) responsible for creating the heaviest elements in our Universe. With larger samples the differences and similarities in mass and history of the dwarf galaxies along with the abundance patterns of the $r$-process enhanced stars themselves can help constrain the astrophysical site of the $r$-process.

\acknowledgements{This publication is based upon work supported by the
  National Science Foundation under grant AST-1108811. This research has made use of NASA's Astrophysics Data System Bibliographic Services. This work has made use of the VALD database, operated at Uppsala University, the Institute of Astronomy RAS in Moscow, and the University of Vienna.

Funding for the DES Projects has been provided by the U.S. Department of Energy, the U.S. National Science Foundation, the Ministry of Science and Education of Spain, the Science and Technology Facilities Council of the United Kingdom, the Higher Education Funding Council for England, the National Center for Supercomputing Applications at the University of Illinois at Urbana-Champaign, the Kavli Institute of Cosmological Physics at the University of Chicago, the Center for Cosmology and Astro-Particle Physics at the Ohio State University, the Mitchell Institute for Fundamental Physics and Astronomy at Texas A\&M University, Financiadora de Estudos e Projetos, 
Funda{\c c}{\~a}o Carlos Chagas Filho de Amparo {\`a} Pesquisa do Estado do Rio de Janeiro, Conselho Nacional de Desenvolvimento Cient{\'i}fico e Tecnol{\'o}gico and the Minist{\'e}rio da Ci{\^e}ncia, Tecnologia e Inova{\c c}{\~a}o, the Deutsche Forschungsgemeinschaft and the Collaborating Institutions in the Dark Energy Survey. 

The Collaborating Institutions are Argonne National Laboratory, the University of California at Santa Cruz, the University of Cambridge, Centro de Investigaciones Energ{\'e}ticas, Medioambientales y Tecnol{\'o}gicas-Madrid, the University of Chicago, University College London, the DES-Brazil Consortium, the University of Edinburgh, the Eidgen{\"o}ssische Technische Hochschule (ETH) Z{\"u}rich, Fermi National Accelerator Laboratory, the University of Illinois at Urbana-Champaign, the Institut de Ci{\`e}ncies de l'Espai (IEEC/CSIC), the Institut de F{\'i}sica d'Altes Energies, Lawrence Berkeley National Laboratory, the Ludwig-Maximilians Universit{\"a}t M{\"u}nchen and the associated Excellence Cluster Universe, the University of Michigan, the National Optical Astronomy Observatory, the University of Nottingham, The Ohio State University, the University of Pennsylvania, the University of Portsmouth, SLAC National Accelerator Laboratory, Stanford University, the University of Sussex, Texas A\&M University, and the OzDES Membership Consortium.

The DES data management system is supported by the National Science Foundation under Grant Number AST-1138766. The DES participants from Spanish institutions are partially supported by MINECO under grants AYA2012-39559, ESP2013-48274, FPA2013-47986, and Centro de Excelencia Severo Ochoa SEV-2012-0234. Research leading to these results has received funding from the European Research Council under the European Union’s Seventh Framework Programme (FP7/2007-2013) including ERC grant agreements 
 240672, 291329, and 306478.
 
The ARC Centre of Excellence for All-sky Astrophysics (CAASTRO) is a collaboration between The University of Sydney, The Australian National University, The University of Melbourne, Swinburne University of Technology, The University of Queensland, The University of Western Australia and Curtin University. CAASTRO is funded under the Australian Research Council (ARC) Centre of Excellence program, with additional funding from the seven participating universities and from the NSW State Government’s Science Leveraging Fund.

This paper has gone through internal review by the DES collaboration.

The authors thank Ian Roederer for helpful discussions and the referee for helpful comments.

}
{\it Facilities:} \facility{Magellan:II (MIKE)}



\begin{thebibliography}{}

\bibitem[Aoki et al.(2007)]{aoki2007} Aoki, W., Beers, T.~C., Christlieb, N.,
  et al., 2007, \apj, 655, 492


\bibitem[Arcones et al.(2007)]{arcones07} Arcones, A., Janka, H.~D., \&
  Scheck, L.\ 2007, A\&A, 467, 1227 

\bibitem[Arcones \& Thielemann(2013)]{arcones2013} Arcones, A., \& Thielemann, F,~K., 2013, Journal of Physics G Nuclear Physics, 40, 1

\bibitem[Asplund et al.(2009)]{asplund2009} Asplund, M., Grevesse, N.,
  Sauval, A. J., \& Scott, P., 2009, \araa, 47, 481


\bibitem[Barklem et al.(2005)]{barklem2005} Barklem, P.~S., Christlieb, N.,
  Beers, T.~C., et al., 2005, \aap, 439, 129

\bibitem[Bauswein et al.(2013)]{bauswein13} Bauswein, A., Goriely, S., \&
  Janka, H.~T. 2013, ApJ, 773, 78 

\bibitem[Bechtol et al.(2015)]{bechtol2015} Bechtol, K.,
  Drlica-Wagner, A., Balbinot, E., et al.\ 2015, \apj, 807, 50

\bibitem[Beers \& Christlieb(2005)]{beers2005} Beers, T. C., \&
  Christlieb, N., 2005, \araa, 43, 531

\bibitem[Bernstein et al.(2003)]{bernstein2003} Bernstein, R.,
  Shectman, S.~A., Gunnels, S.~M., Mochnacki, S., \& Athey,
  A.~E.\ 2003, \procspie, 4841, 1694

\bibitem[Bressan et al.(2012)]{bressan2012} Bressan, A., Marigo, P., Girardi, L., et al. \ 2012, \mnras, 427, 127

\bibitem[Cameron(2003)]{cameron2003} Cameron, A.~G.~W., 2003, \apj, 587, 327 

\bibitem[Carney et al.(2003)]{carney2003} Carney, B.~W., Latham, D.~W., Stefanik, R.~P., et al., 2003, \aj, 125, 293

\bibitem[Castelli \& Kurucz(2003)]{castelli2003} Castelli, F. \& Kurucz,
  R. L., 2003, IAU Symposium, 210

\bibitem[Chubak et al.(2012)]{chubak12} Chubak, C., Marcy, G., Fischer, D.~A., et al.\ 2012, arXiv:1207.6212 

\bibitem[Cohen \& Huang(2009)]{cohen2009} Cohen, J.~G. \& Huang, W., 2009,
  \apj, 701, 1053

\bibitem[Cohen \& Huang(2010)]{cohen2010} Cohen, J.~G. \& Huang, W., 2010,
  \apj, 719, 931

\bibitem[The Dark Energy Survey Collaboration et al.(2016)]{des2016} The Dark Energy Survey Collaboration et al.\ 2016, \mnras, 460, 1270

\bibitem[Dotter et al.(2008)]{dotter2008} Dotter, A., Chaboyer, B., Jevremovi{\'c}, D., et al. \ 2008, \apjs, 178, 89

\bibitem[Drlica-Wagner et al.(2015)]{drlica-wagner2015} Drlica-Wagner,
  A., Bechtol, K., Rykoff, E.~S., et al.\ 2015, \apj, 813, 109

\bibitem[Flaugher et al.(2015)]{flaugher2015} Flaugher, B., Diehl, H.~T., Honscheid, K., et al.\ 2015, \aj, 150, 150


\bibitem[Fran{\c c}ois et al.(2016)]{francois2016} Fran{\c c}ois, P., Monaco, L., Bonifacio, P., et al.\ 2016, \aap, 588, A7 

\bibitem[Frebel et al.(2013)]{frebel2013} Frebel, A., Casey, A.~R., Jacobsen,
  H.~R., \& Yo, Q., 2013, \apj, 769, 57

\bibitem[Freiburghaus et al.(1999)]{freiburghaus99} Freiburghaus, C., Rosswog, S., \& Thielemann, F.~K.\ 1999, ApJL, 525, L121

\bibitem[Fuhr \& Wiese(2009)]{fuhr2009} Fuhr, J.~R. \& Wiese, W.~L., 2009,
  Atomic Transition Probabilities, published in the CRC Handbook of Chemistry
  and Physics, 90th Edition, ed. Lide, D.~R., CRC Press, Inc., Boca Raton, FL,
  10-93 

\bibitem[Fujimoto et al.(2008)]{fujimoto08} Fujimoto, S.~I., Nishimura, N., \&
  Hashimoto, M.~A.\ 2008, ApJ, 680, 1350 

\bibitem[Fulbright et al.(2004)]{fulbright2004} Fulbright, J.~P., Rich, R.~M., \& Castro, S. \ 2004, \apj, 612, 447

\bibitem[Geisler et al.(2005)]{geisler2005} Geisler, D., Smith, V.~V.,
  Wallerstein, G., Gonzales, G., \& Charbonnel, C., 2005, \aj, 129, 1428 

\bibitem[Goriely et al.(2011)]{goriely11} Goriely, S., Bauswein, A., \& Janka, H.~T.\ 2011, ApJL, 738, L32

\bibitem[Gratton et al.(2004)]{gratton2004} Gratton, R., Sneden, C., \& Carretta, E. \ 2004, \araa, 42, 385

\bibitem[Griffen et al.(2016)]{griffen16} Griffen, B.~F., Dooley, G.~A., Ji, A.~P., et al.\ 2016, submitted to \mnras\ (arXiv:1611.00759)

\bibitem[Hansen et al.(2015)]{hansen2015} Hansen T., Andersen, J.,
  Nordstr{\"o}m, B., Beers, T.~C., Buchhave, L.~A. \& Yoon, J., 2015, \aap,
  583, A49 

\bibitem[Hansen et al.(2011)]{hansen2011} Hansen T., Andersen, J.,
  Nordstr{\"o}m, B., Buchhave, L. A., \& Beers, T. C., 2011, \apjl, 743, L1

\bibitem[Hill et al.(2002)]{hill2002} Hill, V., Plez, B., Cayrel, R., et al., 2002, \aap, 387, 560

\bibitem[Jablonka et al.(2015)]{jablonka2015} Jablonka, P., North, P.,
  Mashonkina., et al., 2015, \aap, 583, A67 

\bibitem[Ji et al.(2016a)]{ji2016} Ji, A.~P., Frebel, A., Chiti, A. \& Simon,
  J.~D., 2016a, \nat, 531, 610

\bibitem[Ji et al.(2016b)]{ji2016b} Ji, A.~P., Frebel, A., Simon, J.~D., \& Chiti, A.\ 2016b, \apj, 830, 93 

\bibitem[Johnson et al.(2013)]{johnson2013} Johnson, C.~I., McWilliam, A., \& Rich, R.~M., 2013, \apjl, 775, L27

\bibitem[Kelson(2003)]{kelson2003} Kelson, D.~D.\ 2003, \pasp, 115,
  688 

\bibitem[Kirby \& Cohen(2012)]{kirby2012} Kirby, E.~N. \& Cohen, J.~G., 2012,
  \aj, 144, 168

\bibitem[Komiya \& Shigeyama(2016)]{komiya2016} Komiya, Y., \& Shigemeyama,
  T., 2016, \apj, 830, 76 

\bibitem[Koposov et al.(2015)]{koposov2015} Koposov, S.~E., Belokurov,
  V., Torrealba, G., \& Evans, N.~W.\ 2015, \apj, 805, 130

\bibitem[Korobkin et al.(2012)]{korobkin12} Korobkin, O., Rosswog, S., Arcones, A., \& Winteler, C.\ 2012, MNRAS, 426, 1940

\bibitem[Lattimer \& Schramm(1974)]{lattimer74} Lattimer, J.~M. \& Schramm, D.~N. 1974,\ ApJL, 192, L145

\bibitem[Lemasle et al.(2014)]{lemasle2014} Lemasle, B., de Boer, T.~J.~L.,
  Hill, V., et al., 2014, \aap, 572, A88

\bibitem[Letarte et al.(2010)]{letarte2010} Letarte, B., Hill, V., Tolstoy,
  E., et al., 2010, \aap, 523, A17

\bibitem[Masseron et al.(2014)]{masseron2014} Masseron, T., Plez, B., Van Eck,
  S., et al., 2014, \aap, 571, A47

\bibitem[McWilliam et al.(1995)]{mcwilliam1995} McWilliam, A., Preston, G.~W.,
  Sneden, C., \& Searle, L., 1995, \aj, 109, 2757

\bibitem[McWilliam et al.(2013)]{mcwilliam2013} McWilliam, A., Wallerstein, G., \& Mottini, M., 2013, \apj, 778, 149

\bibitem[Metzger et al.(2010)]{metzger10} Metzger, B.~D., Mart{\'{\i}}nez-Pinedo, G., Darbha, S., et al.\ 2010, \mnras, 406, 2650 

\bibitem[Meyer(1989)]{meyer89} Meyer, B.~S. 1989,\ ApJ, 343, 254

\bibitem[Pe{\~n}arrubia et al.(2008)]{penarrubia2008} Pe{\~n}arrubia,
  J., Navarro, J.~F., \& McConnachie, A.~W.\ 2008, \apj, 673, 226-240

\bibitem[Placco et al.(2014)]{placco2014} Placco, V.~M., Frebel, A., Beers,
  T.~C., \& Stancliffe, R.~J., 2014, \apj, 797, 21

\bibitem[Qian \& Wasserburg(2001)]{QW01} Qian, Y.-Z. \& Wasserburg, G. 2001, \apjl, 552, L55

\bibitem[Ram{\'{\i}}rez et al.(2013)]{ramirez2013} Ram{\'{\i}}rez, I., Allende Prieto, C., \& Lambert, D.~L., 2013, \apj, 764, 78

\bibitem[Roederer et al.(2014b)]{roederer2014b} Roederer I. U., Cowan, J.~J., Preston,  G. W., Thompson, et al., 2014b, \mnras, 445, 2970

\bibitem[Roederer et al.(2011)]{roederer2011} Roederer, I.~U., Marino, A.~F., \& Sneden, C. \ 2011, \apj, 742, 37

\bibitem[Roederer et al.(2016)]{roederer16} Roederer, I.~U., Mateo,
  M., Bailey, J.~I., III, et al.\ 2016, \aj, 151, 82

\bibitem[Roederer et al.(2014a)]{roederer2014a} Roederer I. U., Preston,
  G. W., Thompson, I.~B., et al., 2014, \aj, 147, 136

\bibitem[Rosswog et al.(2014)]{rosswog14} Rosswog, S., Korobkin, O., Arcones, A., Thielemann, F.~K., \& Piran, T.\ 2014, MNRAS, 439, 744

\bibitem[Ryabchikova et al.(2015)]{ryabchikova2015} Ryabchikova, T., Piskunov,
  N., Kurucz., et al., 2015, \physscr, 90, 5

\bibitem[Sadakane et al.(2004)]{sadakane2004} Sadakane, K., Arimoto, N.,
  Ikuta, C., Aoki, W., Jablonka, P. \& Tajitsu, A., 2004, \pasj, 56, 1041

\bibitem[Shetrone et al.(2001)]{shetrone2001} Shetrone, M.~D., C{\^o}t{\'e},
  P. \& Sargent, W.~L.~W., 2001, \apj, 548, 592

\bibitem[Shetrone et al.(2003)]{shetrone2003} Shetrone, M.~D., Venn,
  K.~A. Tolstoy, E., Primas, F., Hill, V., \& Kaufer, A., 2003, \aj, 125, 684

\bibitem[Simmerer et al.(2004)]{simmerer2004} Simmerer, J., Sneden, C., Cowan,
  J.~J., et al., 2004, \apj, 617, 1091

\bibitem[Simon et al.(2015)]{simon2015} Simon, J.~D., Jacobson, H.~R., Frebel,
  A., et al., 2015, \apj, 802, 93

\bibitem[Simon et al.(2016)]{simon2016} Simon, J.~D., Li, T.~S., Drlica-Wagner, A., et al.\ 2016, \apj\, in press (arXiv:1610.05301)

\bibitem[Siqueira Mello et al.(2014)]{siqueira2014} Siqueira Mello, C., Hill,
  V., Barbuy, B., et al., 2014, \aap, 565, A93

\bibitem[Siqueira Mello et al.(2013)]{siqueira2013} Siqueira Mello, C., Spite,
  M., Barbuy, B., et al., 2013, \aap, 550, A122

\bibitem[Sk{\'u}lad{\'o}ttir et al.(2015)]{skuladottir2015}
  Sk{\'u}lad{\'o}ttir, {\'A}., Tolstoy, E., Salvadori, S., et al., 2015, \aap,
  574, A129

\bibitem[Sneden(1973)]{sneden1973} Sneden, C., 1973, \apj, 184, 839

\bibitem[Sneden et al.(2008)]{sneden2008} Sneden, C., Cowan, J.~J. \& Gallino,
  R., 2008, \araa, 46, 241

\bibitem[Sneden et al.(2003)]{sneden2003} Sneden, C., Cowan, J.~J., Lawler,
  J.~E., et al., 2003, \apj, 591, 936

\bibitem[Sobeck et al.(2011)]{sobeck2011} Sobeck, J. S., Kraft, R. P., Sneden,
  C., et al., 2011, \aj, 141, 175

\bibitem[Tafelmeyer et al.(2010)]{tafelmeyer2010} Tafelmeyer, M., Jablonka,
  P., Hill, V., et al., 2010, \aap, 524, A58

\bibitem[Tsujimoto et al.(2015)]{tsujimoto2015} Tsujimoto, T.,
  Ishigaki, M.~N., Shigeyama, T., \& Aoki, W.\ 2015, \pasj, 67, L3

\bibitem[Tsujimoto \& Shigeyama(2014)]{tsujimoto2014} Tsujimoto, T., \&
  Shigeyama, T., 2014, \aap, 565, L5

\bibitem[Venn et al.(2012)]{venn2012} Venn, K.~A., Shetrone, M.~D., Irwin,
  M.~J., et al., 2012, \apj, 751, 102
  
\bibitem[Wanajo et al.(2014)]{wanajo2014} Wanajo, S., Sekiguchi, Y., Nishimura, N., 2014, \apjl, 789, L39

\bibitem[Winteler et al.(2012)]{winteler2012} Winteler, C., K{\"a}ppeli, R., Perego, A., et al., 2012 \apjl, 750, L22

\bibitem[Yong et al.(2013)]{yong2013} Yong, D., Mel{\'e}ndez, J., Grundahl, F., et al. \ 2013, \mnras, 434, 3542

\end{thebibliography}
\end{document}